\documentclass[10pt,twocolumn,aps,prl,superscriptaddress,floatfix, groupedaddress]{revtex4-2}

\pdfoutput=1
\usepackage{amsmath,amsthm,amssymb}
\usepackage[utf8]{inputenc}
\usepackage{mathtext}
\usepackage{times}
\usepackage{epsfig}
\usepackage{amsfonts}
\usepackage{amsmath}
\usepackage{amssymb}
\usepackage{color}
\usepackage{multirow}
\usepackage[normalem]{ulem}
\usepackage{latexsym}
\usepackage{amsfonts}
\usepackage{mathrsfs}
\usepackage{natbib}
\usepackage{verbatim}
\usepackage{graphicx}
\usepackage{xcolor}

\begin{document}
	
	\title{Spatial properties of entangled two-photon absorption}
	\author{D.~Tabakaev, A.~Djorovi\'{c}, L.~La Volpe, G. Gaulier, S. Ghosh, L.~Bonacina, J.-P.~Wolf, H.~Zbinden, R.~T.~Thew}
	\affiliation{D\'epartement de Physique Appliqu\'ee, Universit\'e de Gen\`eve, 1211 Gen\`eve, Switzerland}
	
	\begin{abstract} 
		We experimentally study entangled two-photon absorption in Rhodamine 6G as a function of the spatial properties of a high flux of broadband entangled photon pairs. We first demonstrate a key signature dependence of the entangled two-photon absorption rate on the type of entangled pair flux attenuation: linear, when the laser pump power is attenuated, and quadratic, when the pair flux itself experiences linear loss. We then perform a fluorescence-based Z-scan measurement to study the influence of beam waist size on the entangled two-photon absorption process and compare this to classical single- and  two-photon absorption processes. We demonstrate that the entangled two-photon absorption shares a beam waist dependence similar to that of classical two-photon absorption. This result presents an additional argument for the wide range of contrasting values of quoted entangled two-photon absorption cross-sections of dyes in literature.
	\end{abstract}
	
	\maketitle

	\paragraph{Introduction ---} Two-photon excitation microscopy and spectroscopy techniques are broadly required in both fundamental research and applications due to the relatively high penetration depths and the possibility of 3D slicing~\cite{szoke2020entangled, garcia2020high}. However, these techniques also suffer from fundamental disadvantages, such as low absorption cross-sections~\cite{bebb1966multiphoton,mollow1968two,sperber1986s}. The latter is due to the quadratic dependence of an absorption rate on the photon flux, which is typically compensated by the use of pulsed lasers.
	Theory predicts~\cite{fei1997entanglement, dayan2007theory, schlawin2017entangled, dorfman2016nonlinear} that entangled two-photon absorption (ETPA) is capable of mitigating the small absorption cross-section problem as the photon pairs behave as single quantum objects, which results in a linear rate dependence on the input photon-pair flux, leading to much lower excitation fluxes to obtain the signal. This feature has been observed by several groups experimentally~\cite{lee2006entangled, harpham2009thiophene, upton2013optically, villabona2017entangled, varnavski2017entangled,villabona2020measurements, tabakaev2021energy}, but only the photon pair rate incident to the sample was controlled and varied in previous studies.

	According to theory \cite{fei1997entanglement}, the TPA rate under continuous-wave laser excitation rate $R_{laser}$ [s$^{-1}$] can be written as:
	
	\begin{equation}
		\label{eq:R_TPA}
		R_\text{TPA} = C\,A\,l\,\delta\, \frac{R_{laser}^2}{A^2}\,.
	\end{equation}
	
	\noindent where, $C \, [\text{cm$^{-3}$}]$ is the concentration, $l\, [\text{cm}]$ the sample length, $A\, [\text{cm}^2]$ the beam waist area and $\delta\, [\text{GM}]$ the TPA cross-section. It is clear that the resulting TPA rate scales as $\frac{1}{A}$. Following the same logic, the expression for ETPA rate under continuous-wave photon pair excitation rate $R_{pair}$ [pairs/s] reads as \cite{tabakaev2021energy}:
	
	\begin{equation}
		\label{eq:R_ETPA}
		R_\text{ETPA} = C\,A\,l\,\frac{\delta}{A_eT} \frac{R_{pair}}{A}\,.
	\end{equation}
	
	\noindent where $A_e$ [cm$^2$] is the so-called "entanglement area"  \cite{jost1998spatial} and $T$ is the coherence time of a pair, such that $\sigma_e=\frac{\delta}{A_eT}$ is the ETPA cross-section. Entanglement area is the surface within which a photon of the pair can be found, defined by the uncertainty of its production position and angle. The diameter of this area is defined by the transverse coherence function of the pairs  \cite{schneeloch2016introduction}. In the experimentally relevant case, multiple pairs are produced by spontaneous downconversion (SPDC), and then focused to the sample consisting of multiple molecules or atoms. Typically, the focal spot size is much smaller than the FWHM of the transverse coherence function of the pairs, which means that much more than 50\% of pairs are capable of producing a "coincidence" -- or a two-photon absorption event -- within the beam focal spot size. Under these conditions it is fair to assume that $A_e = A$ and that the ETPA rate depends on the area of the beam waist in the same manner as the TPA rate.

	The conventional way of studying nonlinear optical properties, i.e. absorption and refraction, of a sample, is to use the Z-scan technique~\cite{sheik1990sensitive}. In a classical Z-scan measurement, a thin sample is placed on a translation stage and exposed to a focused laser beam, whose photons can be absorbed in a non-linear process only. Information about the sample properties is derived from the laser power dependence on the translation stage position, i.e. dependence of laser beam losses as the focal spot size of the beam is varied. This method can be used to assess ETPA as well, but instead of laser power, the photon pair coincidence detection rate is needed, which is conventionally obtained using time-correlated single photon counting techniques. 
	However, this is extremely challenging and misalignment of the entangled photon pair beam  due to the sample movement can result in variations of coupling efficiency of the photons after passing through the sample and before the detectors, resulting in a complicated problem of distinguishing coupling losses from pair absorption events~\cite{villabona2017entangled, villabona2020measurements, corona2021experimental}. 
	
	In this letter, we investigate the spatial properties of ETPA by comparing the resultant fluorescence signals in an epi-configuration with those induced by SPA and TPA. We conclude with a discussion on the implication for our understanding of ETPA and its applications in microscopy and spectroscopy.

	\begin{figure}
		\centering
		
		\includegraphics[width=1\linewidth]{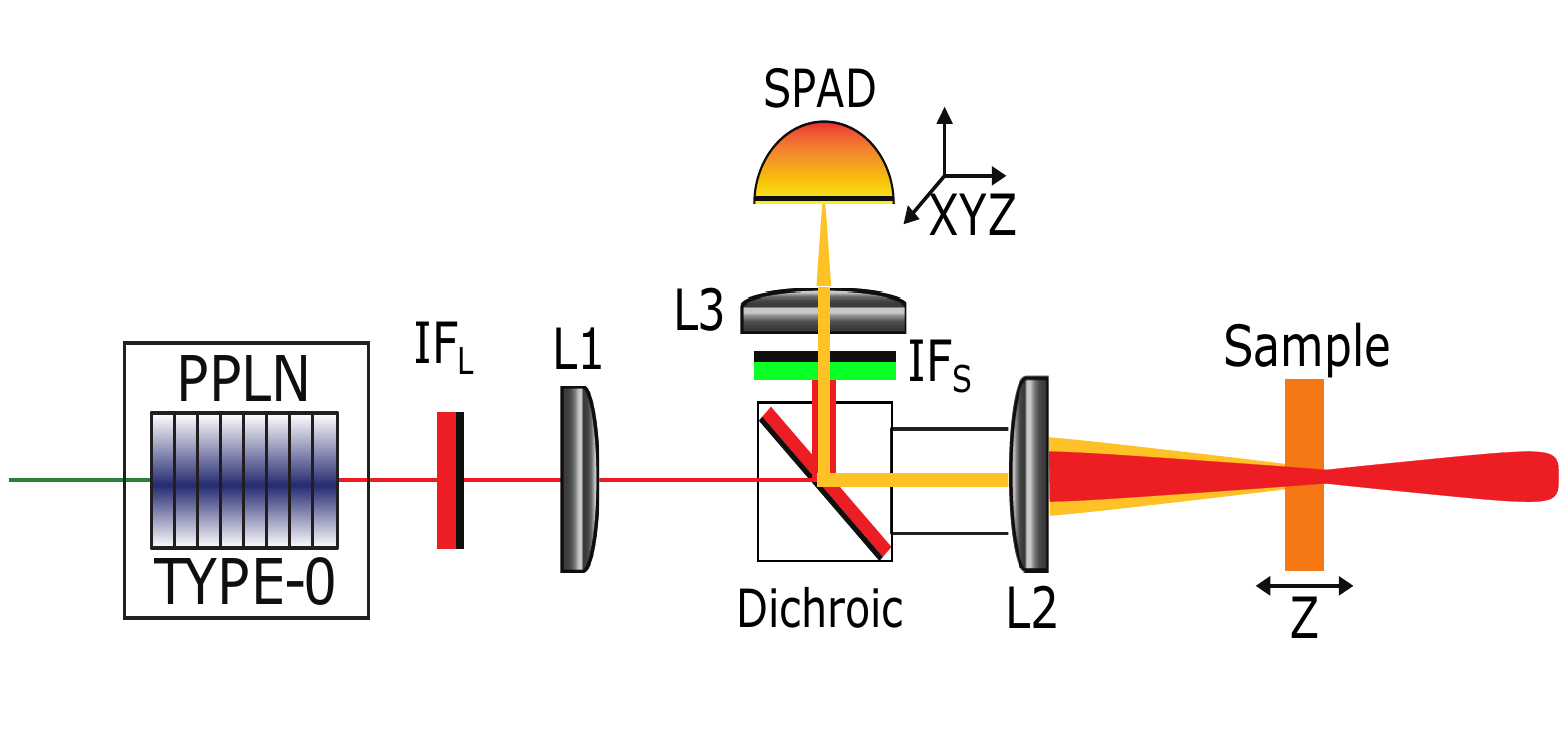}
		\caption{Epifluorescence setup schematic: PPLN -- periodically-poled lithium niobate crystal, pumped by 532\,nm laser; $\text{IF}_{\text{L}}$ -- set of long-pass interference filters; L1 -- pair collimating lens; Dichroic -- dichroic mirror, transparent to IR and reflective to visible light; L2 -- pair focusing and fluorescence collecting lens; Sample -- cell with liquid Rh6G solution; Z -- translation stage; $\text{IF}_{\text{S}}$ -- short-pass interference filter; L3 - fluorescence focusing lens; SPAD - single-photon avalanche diode.}
	\label{fig:Schematic_wide_orthogonal} 
\end{figure}

\paragraph{Experimental set-up ---}Fig.~\ref{fig:Schematic_wide_orthogonal} shows a schematic of the experimental setup. The photon pairs are generated by pumping a 2\,cm periodically-poled Lithium niobate (PPLN) crystal (Covesion MSHG1064-0.5-20, 0.5 x 0.5 mm$^2$ aperture) with 2.5\,W from a 532\,nm continuous-wave laser (Coherent Verdi V5) focused down to a 70\,$\mu$m beam waist by 200-mm (Thorlabs LBF-254-200-A) and 40-mm (Thorlabs LBF-254-040-A) lenses. The laser power is controlled by a 10-cm Glan-Taylor \\Polarizer (Thorlabs GT-10-A) and half-wave plate. The PPLN crystal is temperature phase-matched to produce Type-0 degenerate SPDC pairs with a bandwidth of about 30\,nm, centred at 1064\,nm. The 532\,nm pump laser is blocked by three long pass interference filters ($\text{IF}_{\text{L}}$) (Thorlabs FELH0750, FELH0900 and FELH1050). The photon pair source was characterized similarly to \cite{tabakaev2021energy}: photon pairs were coupled to a single-mode fiber beamsplitter and sent to two single-photon detectors (ID Quantique ID201 and ID220), connected to a time-to-digital converter. By performing time-correlated single-photon counting we found the number of coincidence detections per mW of pump power. Scaling this value to higher pump power and measuring the SPDC beam power with a powermeter (Thorlabs S120C) and the same set of long-pass filters we verified that it scales linearly with the laser pump power. The maximum SPDC power was about 0.16\,$\mu$W ($\sim$\,8.7$\times10^{11}\,\text{s}^{-1}$).

SPDC pairs are collimated by a 10\,cm lens (L1, Thorlabs LBF254-100-C) and, after passing the dichroic mirror (Thorlabs DMLP650R), pairs are focused to the sample mounted on the translation stage (Thorlabs MTS25/M) by a 3\,mm lens (L2, Thorlabs C330TMD-A). The fluorescence from the sample is collected by the same lens L2, before being reflected by a dichroic mirror through three short-pass filters (Thorlabs FESH0650) and then focused by an 11\,mm lens (L3, Thorlabs A397TM-A) to a single-photon avalanche diode (SPAD, ID Quantique ID120, $\sim$ 200 dark counts s$^{-1}$) mounted on a three-axis translation stage.

\paragraph{ETPA signature ---} While ETPA is more efficient with respect to photon flux than TPA, the signal-to-noise ratio (SNR) of experiments relying on entangled photon sources is typically low. 
It is therefore important to ensure that the detected signals are indeed produced by the ETPA process and not arising from, pump leakage, hot-band single-photon absorption \cite{mikhaylov2022hotband} or any other single-photon process. The risk of this misattribution  comes from the indistinguishability of fluorescence signals, produced by these events that would have the same type of (linear) dependence on the input flux. 

\begin{figure}
	\centering
	\includegraphics[width=1\linewidth]{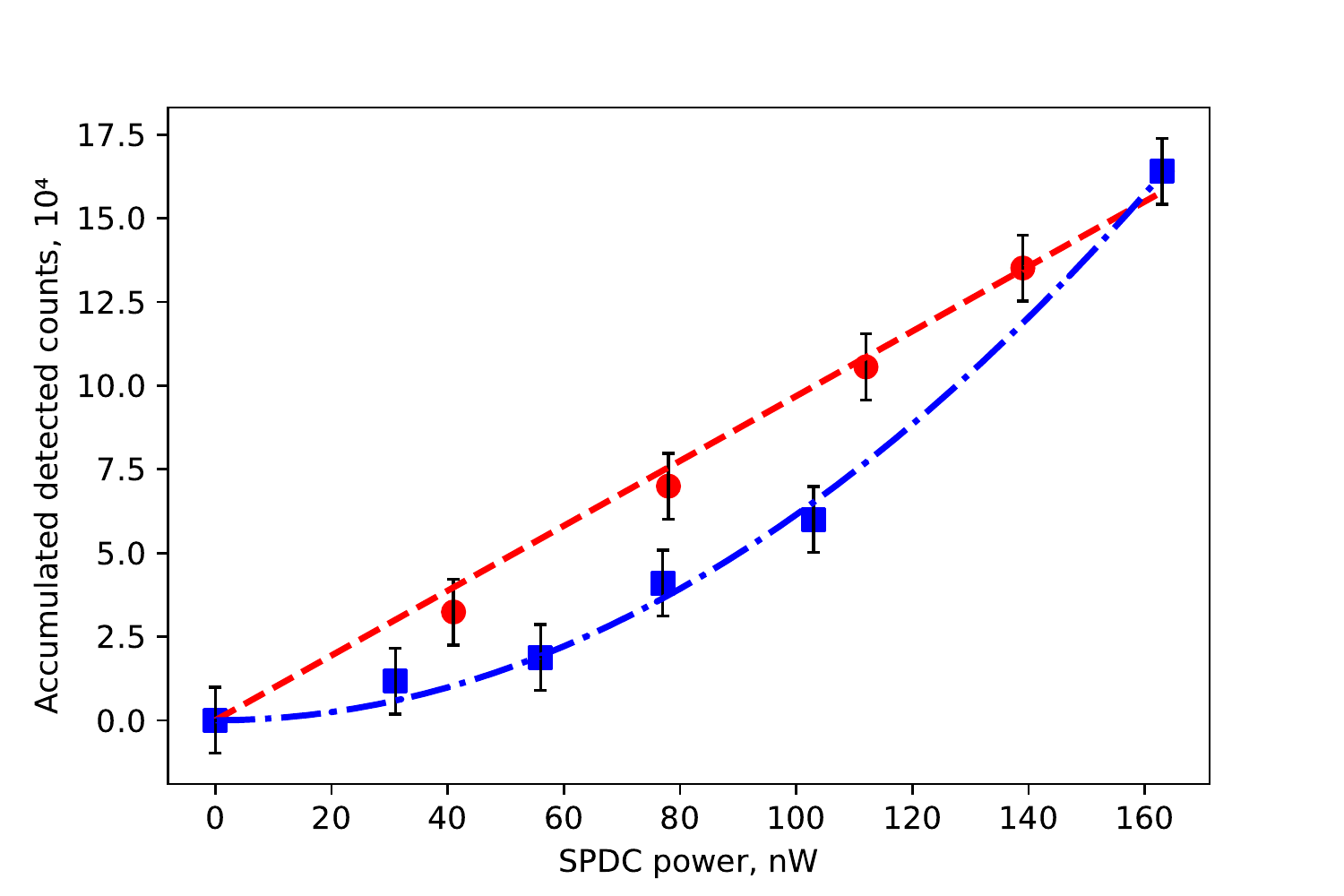}
	\caption{ETPA-induced fluorescence counts as a function of laser pump power attenuation (red circles) and SPDC flux attenuation (blue squares), concentration of Rh6G in ethanol is 5\,mM. Each point is an integration of detector counts over 2$\times10^4$ s, 4.3$\times10^6$ accumulated dark counts subtracted. Error bars are standard deviations over the set of measurements. The red dashed line corresponds to a linear fit and the blue dash-dot line to a quadratic fit of the experimental data.}
	\label{fig:QuadraticLinear}
\end{figure}

We can distinguish these two possible contributions by comparing the fluorescence count rates while attenuating the pump or the photon pair fluxes \cite{dayan2005nonlinear}.
Using the setup from Fig.~\ref{fig:Schematic_wide_orthogonal}, we focused SPDC pairs into a thin home-built cell with 5\,mM liquid solution of Rh6G in ethanol with a lens of 3\,mm focal length. In the first set of  measurements, we controlled the power of the pump laser with a half-waveplate and Glan-Taylor polarizer while detecting the ETPA-induced Rh6G fluorescence for different \textit{pump} attenuations. In the second set of measurements, we increasingly attenuated the SPDC flux by a set of ND filters, and recorded the signal for different \textit{pair} attenuations.
The same measurements were performed with the laser turned off to define the detector dark count level, contributing to the overall signal. 
The results are shown in Fig.~\ref{fig:QuadraticLinear}. The fitting was performed by a least-squares method, resulting in the coefficient of determination R$^2$ = 0.997 for the linear fit and root mean squared error (RMSE) of 0.12 for the quadratic one. A linear dependence in the case of laser beam attenuation, and quadratic in the case of SPDC beam attenuation, confirmed that the measured signal was due to ETPA and not caused by direct detection of scattered pump, down-converted photons or single-photon absorption of any type \cite{mikhaylov2022hotband}. It also constitutes one of the most robust demonstrations of genuine ETPA~\cite{landes2020experimental, landes2021quantifying, raymer2020twophoton}.

\begin{figure} [t!]
	\centering
	\includegraphics[width=1\linewidth]{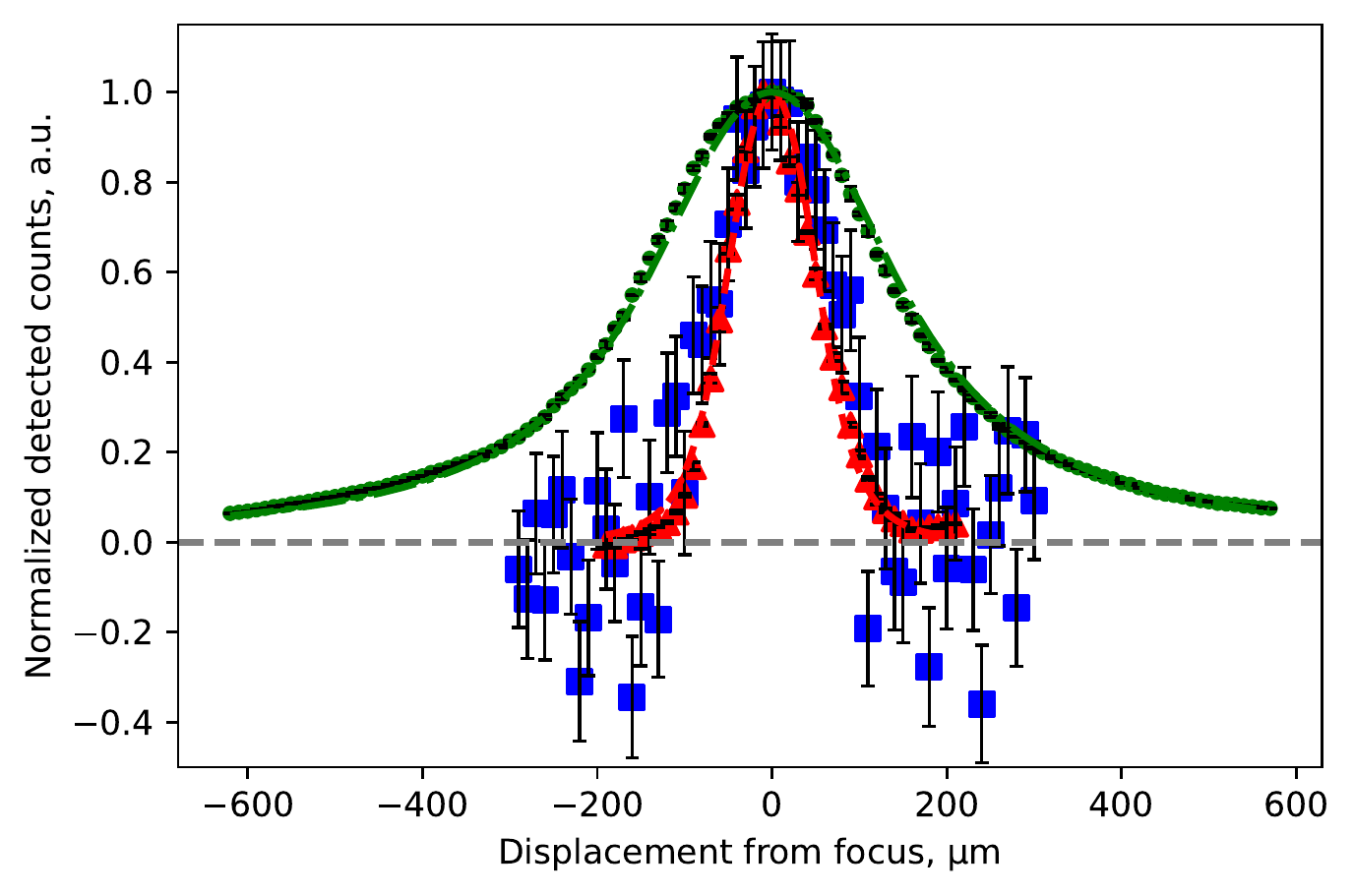}
	\caption{Normalized fluorescence detection rate from a liquid 5 mM Rh6G ethanol solution as a function of translation stage displacement from the focus in the epifluorescence scheme. The fluorescence is induced by: 532\,nm laser photons (green circles); SPDC pairs (blue squares), and 1064\,nm laser photons (red triangles). The SPDC-induced fluorescence data-points are an average of 100 measurements of 1\,s each and the laser-induced are an average of 5 measurements of 1\,s each, measured at 10\,$\mu$m displacement intervals. The red dashed line is the model of the TPA for a Gaussian 1064\,nm laser beam. The green dash-dot line is the model of SPA for a 532\,nm beam. Error bars are Monte Carlo propagated standard deviations. The sizes of error bars for SPA and TPA data points is smaller than the data points.}
	\label{fig:FinalFigure}
\end{figure}

\paragraph{Epifluorescence Z-scan ---} To perform the fluorescence-based Z-scan measurement, the sample is mounted on a translation stage and exposed to the focused SPDC beam. Measuring fluorescence instead of beam attenuation allows us to neglect possible misalignment changing the photon coupling of the entangled pair beam relative to the sample position at the cost of a more challenging alignment of the fluorescence collection optics and single-photon detector position. This choice also increases pump filtering requirements so as to clearly see the fluorescence. Each measurement consists of the integration of fluorescence detections from a home-built cell with a 5 mM liquid Rh6G ethanol solution, which is scanned through the excitation beam's focal point.

In the first instance we use a 1064\,nm continuous-wave laser (Coherent Prometheus) as a source of excitation, which is focused down to a 1.5\,$\mu$m waist to measure classical TPA-induced fluorescence as a reference. Fig.~\ref{fig:FinalFigure} shows the results of this scan where the full width at half maximum (FWHM) of the measured axial profile was about 120\,$\mu$m. This value corresponds well to the 126\,$\mu$m thickness of the sample that was measured using a commercial multiphoton microscope.

\begin{figure}
	\centering
	\includegraphics[width=1\linewidth]{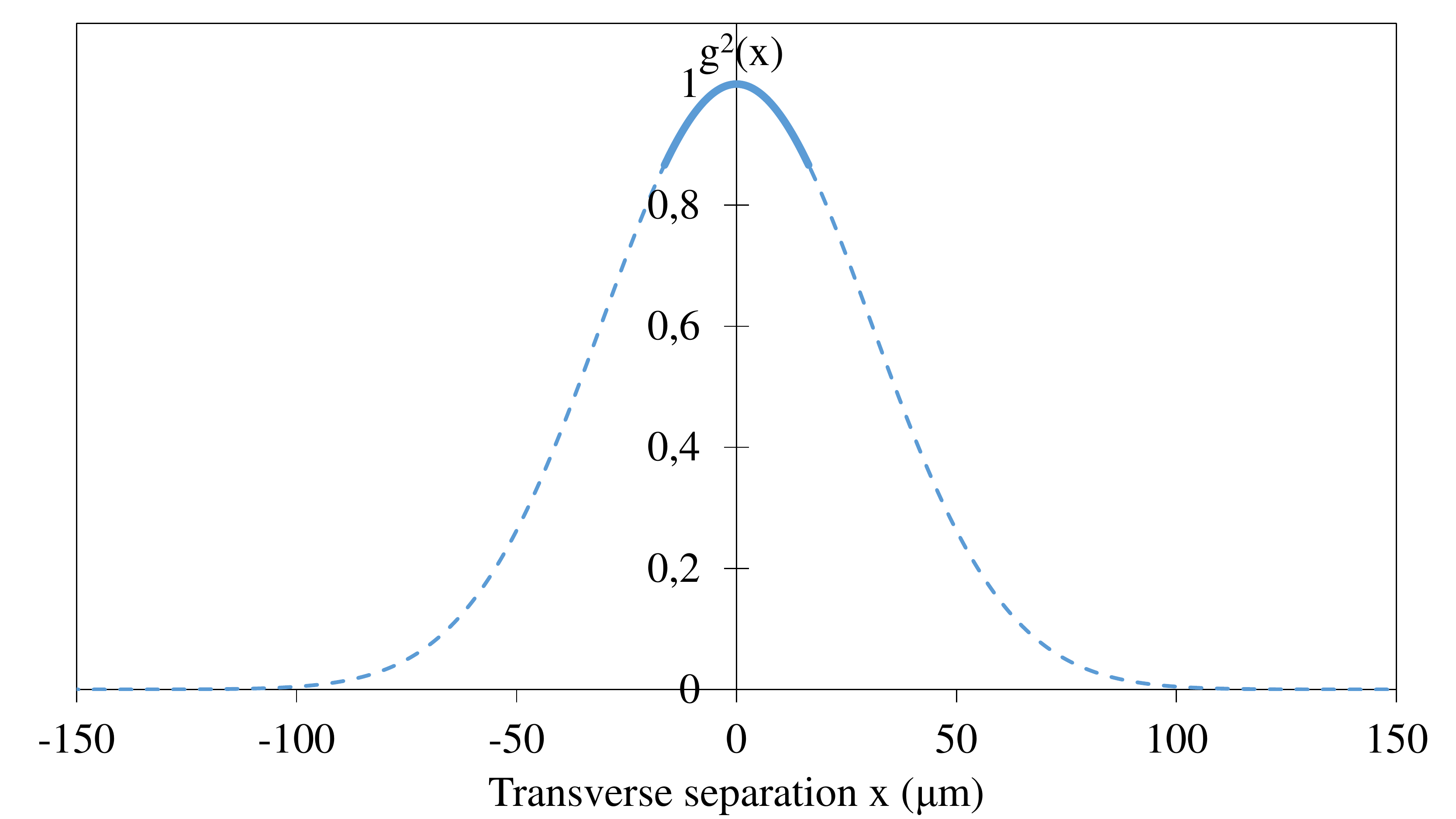}
	\caption{Transverse correlation function of a collinear Type-0 SPDC beam, produced in a 2 cm PPLN crystal, as a function of the transverse separation between the photons of a pair. The thick solid trace on the top represents the maximum separation range of 30\,$\mu$m.}
	\label{fig:g2x} 
\end{figure}

Similarly, we measured the SPA response by focusing a 532\,nm laser down to a 4.4\,$\mu$m waist setting a long-pass filter with cut-off at 550 nm (Thorlabs FELH0550) to ensure only fluorescence photons are detected and any scattered light is blocked.

We finally injected the entangled photon pairs at maximum laser pump power (about 8.7$\times10^{11}\,\text{s}^{-1}$, 0.16\,$\mu$W) and focused them down to a 4.5\,$\mu$m waist on the cell with liquid Rh6G solution. The signal (10 counts/s) obtained by replacing the Rh6G sample by pure ethanol was subtracted for background correction. To ensure that we stay within the assumption of $A_e=A$, we take an extreme case for the maximum beam size during this measurement of 30 $\mu$m as a transverse separation between photons in a pair and along the same lines as in ~\cite{schneeloch2016introduction}, we calculate that the maximum drop of the transverse correlation function is about 10\% as shown on Fig.~\ref{fig:g2x}.

In Fig.~\ref{fig:FinalFigure} we see all three measurement results. It is clear that the FWHM of the SPDC Z-scan profile lies between the widths of the two classical references obtained in the same geometry. To compare the results we first modeled the fluorescence rate which can be detected when using an undepleted Gaussian beam for TPA and SPA taking into account the 1064 nm and 532 nm laser beam parameters respectively~\cite{mertz2019introduction}. We start from  calculating effective waists of the beam $w_z$ and $w_{zTP}$ in case of single- and two-photon absorption respectively
\begin{gather}
	\label{eq:wz}
	w_z = \sqrt{\frac{w_0^2+\frac{\lambda^2}{4\pi^4\textit{NA}^2}+2w_\text{d}^2}{\frac{\lambda^2}{(4\pi^4w_0^2)+\textit{NA}^2}}} \\
	w_{zTP} = \sqrt{\frac{w_0^2+\frac{\lambda_{TP}^2}{2\pi^4\textit{NA}^2}+2w_\text{d}^2}{\frac{\lambda_{TP}^2}{(4\pi^4w_0^2)+2\textit{NA}^2}}},
\end{gather}

\begin{figure} [t]
	\centering
	\includegraphics[width=1\linewidth]{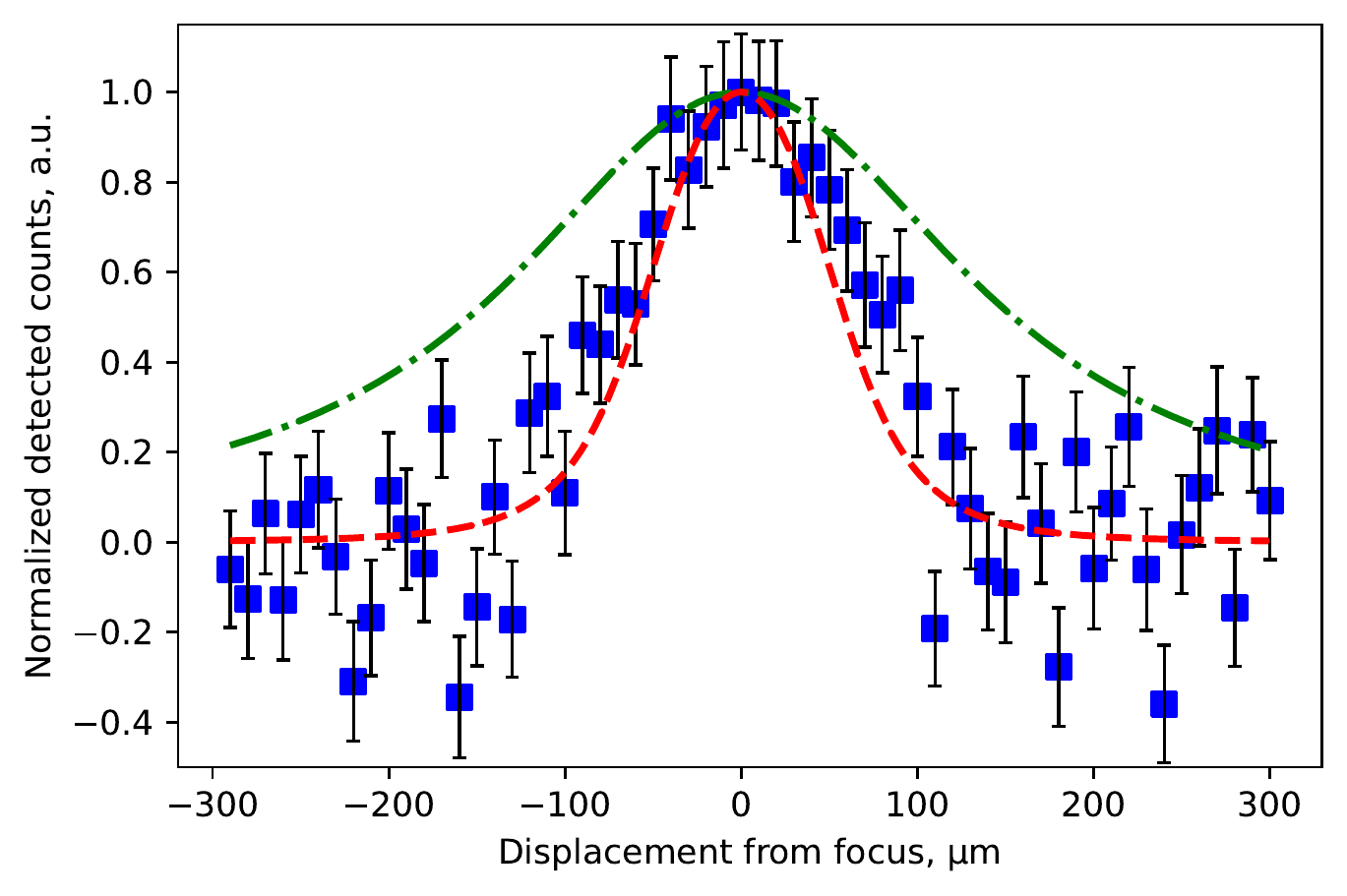}
	\caption{Normalized fluorescence detection rate from a liquid 5 mM Rh6G ethanol solution as a function of translation stage displacement from the focus in the epifluorescence scheme under the SPDC excitation (blue squares), and models of fluorescence rate under single- (red dashed line) and two-photon excitation (green dash-dot line), described in equations \eqref{SPA} and \eqref{TPA}.} \label{fig:ETPA_fit}
\end{figure}

\noindent where $w_0$ is a beam waist size at the focus, $NA=0.7$ is the numerical aperture of the excitation and collection lens L2, $\lambda$ and $\lambda_{TP}$ are excitation wavelengths for single- and two-photon absorption cases respectively, $w_d$ is the detector area (detector diameter is 500 $\mu$m). We then put these values into the expressions describing the normalized fluorescence detection rate for SPA and TPA given by

\begin{align}
	&R^{SPA}(z)=\arctan \left(\frac{z+d}{w_z}\right)-\arctan \left(\frac{z-d}{w_z}\right) \label{SPA} \\
	&R^{TPA}(z)=w_{zTP}(\arctan \left(\frac{z+d}{z_R}\right)-\arctan \left(\frac{z-d}{z_R}\right))\nonumber\\ 
	&-z_R(\arctan \left(\frac{z+d}{w_{zTP}}\right)-\arctan \left(\frac{z-d}{w_{zTP}}\right)) \label{TPA}
\end{align} 

Here $z$ is the displacement of the sample relative to the beam focus, $d$ is the sample thickness and $z_R$ is the Rayleigh range. To fit the experimental data on Fig.~\ref{fig:FinalFigure}, a sample thickness $d$, the detector size and focal spot size $w_0$ were left as free parameters and a least square optimization algorithm used the measured values of these parameters as a starting guess. The resulting fits of the measured fluorescence rates yield a RMSE of 0.017 in the SPA case and 0.035 in the TPA case.

To fit the ETPA data, we fixed the excitation wavelength at the central wavelength of SPDC pairs, at 1064 nm, and used the parameters obtained from the fitting of classical references and the result is demonstrated on the Fig~\ref{fig:ETPA_fit}: when fitted by a model of the fluorescence rate produced under the single-photon excitation RMSE is 0.383, while TPA model fits with RMSE of 0.171.

\section{Discussion}		
In Figure~\ref{fig:QuadraticLinear} we demonstrated one of the clearest signatures of ETPA both highlighting linearity of its rate as a function of SPDC rate and also that experimentally the fluorescence is induced by the pairs and not from experimental artifacts. This provides the baseline to unambiguously study the spatial characteristics in the epi-detection configuration that is of relevance in the context of microscopy.

We modeled the transversal coherence function of our photon pair source and demonstrated analytically that ETPA rate as a function of the beam size scales in the same way as TPA rate. Using the fluorescence-based Z-scan we demonstrated the similarity between shapes and widths of TPA and ETPA spatial profiles. We further reinforced our analysis by modeling the fluorescence rate under single- and two-photon excitation and confirmed that despite the fact that ETPA rate scales linearly as a function of excitation photon pair rate, its spatial properties follow a TPA-like behavior. The spatial dependence of the ETPA rate has broader implications in studies of ETPA processes. Consistent and standardized reporting of ETPA rate values is critical to untangling the many contrasting reported values of ETPA cross-sections. Our results suggest that reporting an ETPA cross section, $\sigma_e*$, value for a given system is insufficient. Due to this dependence on spatial properties, we believe that $\sigma_e*A$, where A is the cross-section of the SPDC beam at focus, is a more pertinent choice of value to use when comparing different optical systems and experiments. These results suggest further investigations are required into the effects of quality of focus, spatial aberration and choice of imaging system on the ETPA rate. For example, the line on the Fig.~\ref{fig:QuadraticLinear} corresponds to $\sigma_e$ on the order of $\sim5*10^{-22}$cm$^2$, which corresponds up to an order of magnitude to previously obtained value for Rh6G study \cite{tabakaev2021energy}, and combined with the beam waist area $\sigma_e*A\approx2*10^{-34}$. However, due to the case of typically low SPDC intensities, ETPA-induced signals and challenging SNR, spatial properties of ETPA call for more studies.

\paragraph{Acknowledgements ---} We acknowledge support from the Swiss National Science Foundation through the Sinergia grant CRSII5-170981. We also wish to thank Ralph Jimenez, Martin Stevens and Michael Raymer for many
fruitful discussions and insights.

\bibliography{Zscan}

\end{document}